\pdfoutput=1
\documentclass[10pt]{article}
\usepackage[letterpaper]{geometry}
\usepackage{hicss51}
\usepackage{times}
\usepackage[none]{hyphenat}
\usepackage[hyphens]{url}
\usepackage{latexsym}
\usepackage{graphicx}
\graphicspath{{images/}}

\usepackage[table]{xcolor}

\usepackage{doi}
\usepackage{indentfirst}

\title{Investigating the Intersection of  Science Fiction, \\ Human-Computer Interaction and Computer Science Research}

\author{Philipp Jordan \\
  University of Hawai`i at M{\=a}noa\\
{\href{mailto:philippj@hawaii.edu}{philippj@hawaii.edu}} \\}

\date{}

\usepackage{hyperref}
 \hypersetup{
     colorlinks=true,
     linkcolor=blue,
     filecolor=magenta,      
     urlcolor=cyan,
 }
\begin{document}
\maketitle
\begin{abstract}
This paper outlines ongoing dissertation research located in the intersection of science fiction, human-computer interaction and computer science. Through an interdisciplinary perspective, drawing from fields such as human-computer interaction, film theory and studies of science and technology, qualitative and quantitative content analysis techniques are used to contextually analyze expressions of science fiction in peer-reviewed computer science research repositories, such as the ACM or IEEE Xplore Digital Libraries. This paper concisely summarizes and introduces the relationship of science fiction and computer science research and presents the research questions, aims and implications in addition to prior work and study methodology. In the latter part of this work-in-progress report, preliminary results, current limitations, future work and a post-dissertation trajectory are outlined.

\end{abstract}

\section{Introduction}
How much of a role does science fiction\footnote{`Science fiction' (SF), as used in this paper, encompasses science fiction literature as well as science fiction movies and shows. Interactive media, such as science fiction-based video games are not considered in the current research framework.} (SF) play in `real-world' scientific innovation \cite{Kirby.2010b} and progress? What are the opportunities and hindrances -- if any -- of SF in the context of computer science research, for example computer science education \cite{Bates.2012,Myers.2016} or artificial intelligence \cite{DBLP:journals/corr/BurtonGKKMW17} (AI) ethics? Should computer scientists actually study \cite{Bausells.2015} SF?  

Despite the fact that discussions of the synergy effects between SF and computing research are a regular topic in popular news and tech magazines, the bi-directional relationship of both domains is in reality not accurately described, nor fully scrutinized. In particular, studies which investigate the presence, nature, and patterns of use of SF in the context of computer science research are scarce and rely mostly scattered oral accounts. Therefore, evidence-based investigations are critical in order to better understand the potential utility and latent shortcomings of SF for future computing and human-computer interaction (HCI) research, innovation and education. In this context, this dissertation endeavors to create a better understanding of the relationship between SF and computer science research.

\section{Research Aims and Objectives}
This research draws from an interdisciplinary perspective and distinct bodies of literature from the following knowledge domains and topical areas:
\begin{enumerate}
    \setlength{\itemsep}{0pt}
    \setlength{\parskip}{0pt}
    \setlength{\parsep}{0pt}   
    \item SF, computing visions and design theory
    \item SF in contemporary HCI research
    \item SF movies and real-world scientists
    \item SF in computer science education
    \item SF and computer science frameworks and models
    \item SF and evolutionary computer science research
\end{enumerate}
The principal research objective of the dissertation is the discovery, description and analysis of the relationship between SF and computer science research through a content analysis of scientific publications, which refer SF in their full-text and metadata; over time and within distinct subfields of computer science research. 

As introduced earlier, there are scattered indicators of the introduced relationship which can be found in the contemporary public news \cite{McDonald.2014,Rifkin.2016}. tech magazines \cite{Anders.2015} and expert communities \cite{NASA.2016,NationalAcademyofSciences.2018}. While present, these traces are typically of anecdotal nature with no clear evidence as in how SF is used in, linked with, or has contributed to, a specific research project, product, or process. While SF in computer science research itself is discussed tangentially on occasion (e.g. \cite{Marcus:1999:OPS:632716.632775,Larson.2008,Kurosu}), proper studies with focus on the subject matter are rare and, if present: 
\begin{itemize}
    \setlength{\itemsep}{0pt}
    \setlength{\parskip}{0pt}
    \setlength{\parsep}{0pt} 
    \item limited on specific aspects and applications of SF for computer science research, for example as data sources \cite{Figueiredo.2015,Troiano.2016};
    \item have a selection bias and lack objective and longitudinal data to describe the relationship of SF and HCI over time and across fields of computer science research (e.g \cite{Larson.2008,Schmitz.2008});
    \item or use a limited sample size and lack focus on SF and computing research (e.g. \cite{Levin.2014a,Levin.2014b}).
\end{itemize}
Currently, no meta-study does investigate the presence and patterns of usage of SF in the context of scientific publications. However, such investigations are important as they can point to missed opportunities and future potentials of SF for computing research. For example, an overview and evolutionary description of SF in computing research can provide important insights for future computing research. Perhaps, data on the occurrence, frequency, usage and contextual referrals of SF and its terminology derivatives in computer science publications can identify emerging computing research themes (which are more or less prone to use SF; and therefore, might benefit more or less from assimilating SF in the future). In addition, the cultural origin, purpose, and nature\footnote{For example, dystopian versus utopian SF.} of the SF narratives in computer science publications have not been described as well, for instance, the patterns of usage of Western and non-Western SF. Also, the prevalence or abundance of SF referrals within and between distinct subfields of computer science research, for instance, the patterns of usage in HCI, human-robot interaction (HRI) or design research are not investigated at present.

One might ask why SF is used in a computer science publication; what kind of SF is referenced in the respective publication and how does it inform the research itself? Is there a cultural bias of the referred SF in computing literature? Did SF referrals evolve from rhetorical devices into serious research topics for HCI and computing literature over time? Are there clusters of computer science areas (e.g. HCI, HRI, AI) which regularly draw from and refer to SF? These are research questions this study aims to answer and it is in this niche, where the contribution of this dissertation is located in.

\paragraph{Research assumptions, aims and objectives}The assumption in this study is that SF is an existing, yet undiscovered topical domain in computer science research and its associated subfields. Specifically, SF movies, but also general SF authors, technologies, concepts, characters, and stories will be traced in order to investigate the prevalence, usage, and appropriation of SF in scientific publications. Thus, the primary aims of this dissertation are to establish a ground truth of the relationship between HCI and SF. 
First, the dissertation aims to quantify, for the first time, the described relationship between popular SF media depiction of technology in computer science publications. Second, the dissertation aims to create and describe measures of this relationship through a quantitative content analysis of publications which reference SF, as well as descriptive, qualitative analysis. Third, the dissertation aims to inform a more comprehensive framework to re-conceptualize the bi-directional relationship of computer science, HCI and SF, outlined through specific research objectives (RO) below. 

\paragraph{RO 1: Identify an interplay between SF and HCI Research}

The primary research objective of this dissertation is to discover a relationship between SF and computer science research, with a special focus on human-computer interaction (HCI) topics. For the purposes of this dissertation, RO1 is addressed through the related and prior work. The related work provides a thematic literature review and indicates a mutual, yet not quantified, relationship of both fields. The prior work   introduces three pilot studies \cite{Jordan2017,Jordan2018,THRI_in_Review} which confirmed the presence and usage of SF concepts in the ACM Digital Library (DL), respectively the CHI proceedings. To further investigate these suggestions and corroborate RO1, an in-depth study of the occurrence of SF in a larger scientific corpus of peer-reviewed publications is conducted in a comparable collection, that is the IEEE \textit{Xplore} DL repository.

\paragraph{RO 2: Determine patterns of SF references in computer science / HCI research publications}
The second research objective of this dissertation is to determine a quantifiable, that is to say, measurable, impact SF has had, currently has and potentially will have on computer science and HCI research. To do so, quantitative text mining in combination with a qualitative content analysis of the referred SF concept(s) in relationship to the research theme of the retrieved publications are applied in order to understand the context of the SF referral. Through such an analysis, a determination of the purpose and level of influence of the referral in specific subtopics of computing research and HCI (such as mobile computing, robotics, wearables, virtual reality and others) can be assessed. Moreover, patterns across periodicals and collections, such as the mechanics of SF referrals in certain subtopics in computing literature could indicate which fields of research or authors are more or less prone to draw from, and integrate, SF in their research.

\paragraph{RO3: Describe and analyze the characteristics, purpose and cultural origin of the SF referenced in computer science / HCI publications over time}
A third research objective in this dissertation is the identification and determination of the cultural origin of the referenced SF in the retrieved publication(s) in order to provide an evolutionary overview of the type and purposes of SF in computer science research over time. A contextual analysis of SF in computer science research publications can describe the cultural origins of the SF material (e.g. authorship, country of production) as well as the purposes of the SF referral in the context of the individual publication. Through RO3, the prevalence and absence of specific SF from specific countries and areas will be determined. Patterns, for instance, the type of SF referral and the purpose of the reference in correlation to the publication year and computer science subfields over time should enable deductions if SF precedes, matches or follows real-world technological development and research. 
\section{Prior Work}
The dissertation topic  has been developed since 2015 and preliminary results have been published on an ongoing basis with collaborators. In 2015, an workshop paper \cite{Jordan2015} was presented at the Workshop on SF and the Reality of HCI: Inspirations, Achievements or a Mismatch in the context of the 27\textsuperscript{th} Australian Conference on Human-Computer Interaction (OzCHI). In 2016, a theoretical paper \cite{Jordan2016} and poster on the conceptualization and measurement of science ficiton in computer science research was presented at the 8\textsuperscript{th} International Conference on Human-Computer Interaction (HCII). As co-author, a short paper \cite{Mubin:2016} on the relationship of SF and scientific inspiration was published at the 13\textsuperscript{th} International Conference on Advances in Computer Entertainment Technology (ACE). In 2017, the first pilot study on Star Trek in the ACM DL was published \cite{Jordan2017}. In 2018, the second pilot study was published in the context of the 2018 HCII conference \cite{Jordan2018}.

\paragraph{Awards, Earned Press and Honors}
The dissertation research and preliminary results have gained attention in the information science and HCI communities. In 2016, an external research grant \cite{Mubin.grant.2016}, through a collaboration with a senior researcher from the University of Western Sydney, Australia, of \$3,500 was awarded. In 2018, the latest paper of the research project \cite{Jordan2018} was referenced in the MIT Emerging Technology News \cite{EmergingTechnologyfromthearXiv.}. An ACM Interactions (IX) blog post \cite{Jordan2018_IX} on the topic of SF and HCI was published in early 2018. Furthermore, the doctoral candidate was awarded an Eugene Garfield Dissertation Fellowship \cite{Jordan_Garfield_Award} in 2018 for the dissertation research, valued at \$3000, which honors doctoral candidates working on their dissertations in the library and information science, information studies, informative, or a related field .
\section{Methodology}
Currently, the final analysis of  one large dataset of 1400 computer science publications, retrieved in the IEEE \textit{Xplore} DL, is prepared for analysis. Specifically, the corpora metadata characteristics and contextual uses, purposes, and types of SF references within each individual publication will be examined. 
Based on the prior work, the proposed methodology will follow a two-stage, combined, qualitative content analysis (bottom-up) and quantitative, text analysis (top-down) approach. Preceding that final study and content analysis, inter-rater agreement and a  coding scheme will be established and defined. The SF referrals and contextual characteristics in the individual publication are will be manually reviewed and coded in the bottom-up approach. In the top-down approach, an assisted, quantitative text analysis of the corpora metadata will be conducted. For the successive analysis, the results of both approaches will be consolidated and compared against each other. Detailed steps of the research methodology are outlined below.

\paragraph{Step 1: Initial searches and candidate sets}
In order to estimate the occurrence of SF and its related concepts, initial exploratory searches for SF and term derivatives of the concept in selected computer science collections, such as the IEEE \textit{Xplore} DL, have yielded multiple candidate sets for potential further investigation and qualitative analyses.
\paragraph{Step 2: Selection and acquisition of publications}
In a second step, a final candidate dataset\footnote{Currently, this dataset is a set of 1400 publications, retrieved via  full-text search for ``science fiction'' in the IEEE \textit{Xplore} DL.}, based on the feasibility of the methodology, is acquired. The final dataset will be first fully downloaded and then imported into content analysis software for a further qualitative inspection with regards to the prior outlined research aims and objectives.  
\paragraph{Step 3: Initial assessment of publications}
In a third step, the final candidate set is explored through a qualitative and guided quantitative coding and text analysis.
The retrieved publications will be assessed with regards to the focus of the research publication itself (e.g. AI, augmented reality, gestural interactions) in order to identify subtopics in computer science research which are more predisposed to reference SF than others. Finally, an evolutionary analysis over time and across periodicals and proceedings of the retrieved publications with a focus on the themes discussed in the year or decade the research was published in should enable a measurement of the foresight, impact and overall usefulness of SF in computer science research. 
\paragraph{Step 4.1: Qualitative coding}
Prior work \cite{Jordan2018} proposes five research themes for a qualitative coding of SF in conjunction with the main research theme of the computer science research publication:
\begin{enumerate}
    \setlength{\itemsep}{0pt}
    \setlength{\parskip}{0pt}
    \setlength{\parsep}{0pt}  
    \item \textbf{Theoretical}:  Publications on design research, ideation,  and design fiction.
    \item \textbf{New interactions:}  Novel interfaces and interaction modes (e.g. gestural, haptic, shape-changing).
    \item \textbf{Human-body modification / extension:} End-of-life technologies, on-body fabrication of artifacts, implants and in-body insertables.
    \item \textbf{Human-Robot Interaction and AI}: Human-robot or human-agent interaction and agency, artificial intelligence, natural language interfaces, and ethics.
    \item \textbf{HCI and Future Visions of Technology}: Technology in conjunction with agency and power, utopian versus dystopian dichotomies in the tension field of ubiquitous computing (e.g. privacy versus security).
\end{enumerate}
\paragraph{Step 4.2: Quantitative analysis} 
Publication metrics (e.g. citations, venue, downloads) as well as document metadata analyses (e.g. term-frequency, co-word or keyword analysis) of the final dataset will be used to identify which topical domains in computing literature are more or less connected with SF\footnote{For such a potential content analysis software, Provalis QDA Miner \cite{ProvalisResearch.2017} provides functionalities for bivariate comparisons of variables, for example, to compare the venue of publication or publication year with any given coded text, such as the purpose of the SF referral.}
\paragraph{Step 5: Consolidation of the qualitative and quantitative analysis} For the successive analysis, the results of both approaches will be consolidated and compared against each other in order to generate best-possible insights on the results, and subsequently derive conclusions to in relationship to the outlined research questions and objectives.

\section{Implications and Future Work}

\paragraph{Implications}
The results of this study are expected to discover novel links and synergy effects of computer science and SF; over time and across categorical subdisciplines of computing research. Furthermore, the evolutionary analysis of the results of this study are expected to demonstrate both, that SF has been a topic of interest since the inception of the field of computer science, and is increasingly applied in present-day, computer science research. The implications of this study are further expected to guide future computer scientists and educators to consciously utilize SF in their teaching, research and scholarship and therefore drive future innovative HCI and computer science research, application, and education. This study also aims to provide a methodological contribution to the field of studies of science, technology and society (STS) in form of a case study by disentangling the underlying, ambiguous relationships between SF and computer science -- or in a broader definition -- science and art.


\paragraph{Future Work}
At the present time, one journal paper on the usage of SF robots in computer- and HRI-research is in review at ACM THRI \cite{THRI_in_Review}. A second manuscript, an analysis of the cultural prevalence of specific science fiction material in the full CHI proceedings is close to submission for an HCI journal. A third paper, which aims to propose science fiction media (especially audio-visual media, such SF movies and shows) as design fictions (see e.g. \cite{Lindley:2015:BFY:2783446.2783592}) for HCI and computer science research is in the conceptual phase with the aim for submission by the end of 2018.

Future studies to extend the presented research trajectory will aim to conduct qualitative interviews with HCI researchers in order to investigate and assess the reasons, and reservations, of researchers and authors in computer science, who refer sci-fi in their research output.


\bibliographystyle{ieeetr}
\bibliography{bib}

\end{document}